# Development of an Equation-based Parallelization Method for Multiphase Particle-in-Cell Simulations

Mino Woo[1,2], Terry Jordan[1], Tarak Nandi[1], Jean François Dietiker[3], Christopher Guenther[1], and Dirk Van Essendelft[1,*]

[1]National Energy Technology Laboratory, Morgantown, WV, 26505, United States
[2]Oak Ridge Institute for Science and Education, Oak Ridge, TN, 37830, United States
[3]Leidos Research Support Team, Pittsburgh, PA, 15236-0940, United States

**Abstract:** Manufacturers have been developing new graphics processing unit (GPU) nodes with large capacity, high bandwidth memory and very high bandwidth intra-node interconnects. This enables moving large amounts of data between GPUs on the same node at low cost. However, small packet bandwidths and latencies have not decreased which makes global dot products expensive. These characteristics favor a new kind of problem decomposition called "equation decomposition" rather than traditional domain decomposition. In this approach, each GPU is assigned one equation set to solve in parallel so that the frequent and expensive dot product synchronization points in traditional distributed linear solvers are eliminated. In exchange, the method involves infrequent movement of state variables over the high bandwidth, intra-node interconnects. To test this theory, our flagship code Multiphase Flow with Interphase eXchanges (MFiX) was ported to TensorFlow. This new product is known as MFiX-AI and can produce near identical results to the original version of MFiX with significant acceleration in multiphase particle-in-cell (MP-PIC) simulations. The performance of a single node with 4 NVIDIA A100s connected over NVLINK 2.0 was shown to be competitive to 1000 CPU cores (25 nodes) on the JOULE 2.0 supercomputer, leading to an energy savings of up to 90%. This is a substantial performance benefit for small- to intermediate-sized problems. This benefit is expected to grow as GPU nodes become more powerful. Further, MFiX-AI is poised to accept native artificial intelligence/machine learning models for further acceleration and development.

**Conflicts of interest:** The authors have no conflicts of interest to declare that are relevant to the content of this article.

---

[*] Corresponding author. E-mail address: dirk.vanessendelft@netl.doe.gov (D. Van Essendelft)

**Article Highlights:**

- An intuitive equation-based parallelization without domain decomposition
- Infrequent movement of large data rather than frequent movement of small data
- The port of Multiphase Flow with Interphase eXchanges (MFiX) to TensorFlow
- Replication of nearly identical results with 4-9 significant digits matching
- Comparable performance of a 4-GPU workstation against 25 CPU nodes (1000 cores)

# 1 Introduction

Recent advances in high-performance computing (HPC), including those in both multi central processing units (CPUs) and graphics processing units (GPUs), have greatly improved the speed of numerical computations in areas like computational fluid dynamics (CFD) [1, 2], molecular dynamics [3, 4], lattice-Boltzmann methods [5], and deep learning [6, 7]. Computationally expensive CFD computations requiring high mesh resolution have significantly benefited from these advances. However, developing parallel computer codes to utilize multicore architecture is still challenging as it requires rewriting and optimization of existing serial codes, particularly when both CPUs and GPUs are used in a collaborative manner [8, 9], where the GPU to CPU memory transfer can be a serious bottleneck [10]. Although parallel computing has been widely used in a variety of CFD applications [11, 12], development of a computer code to fully utilize multicore HPC capabilities on heterogeneous systems is still challenging [13].

TensorFlow (TF) [14] is an open-source software library capable of utilizing both CPU and GPU architectures and can potentially help accelerate CFD computations. TF was the most popular and well-documented framework at the beginning of this project and is an effective tensor algebra library, which makes it useful for many scientific computing challenges. In TF, numerical computations are represented as a graph of connected operations [14] that can be executed on multiple devices including multicore CPUs, general purpose GPUs, and custom-designed application-specific integrated circuits (ASICs). This enables TF-based applications to be run on a wide variety of hardware and to scale well from powerful GPU servers down to mobile devices [15]. In addition, TF offers multiple levels of abstraction and operates on multiple languages and operating systems [14]. These features allow TF to substitute or supplement the existing CFD frameworks to accelerate computations. At its heart, TF is an extremely powerful math library that allows for both hardware agnostic programming and hardware specific optimization depending on the requirements. Recently, Zhao et al. [16] reported that a TF-based CFD simulation with GPU acceleration required almost ten times less computational time compared to the same simulation carried out on CPUs using a Fortran code.

The MFiX (Multiphase Flow with Interphase eXchanges) CFD code was chosen as the base development platform to investigate GPU acceleration using TF. MFiX is an open-source multiphase flow solver developed at the National Energy Technology Laboratory (NETL) and employs the two-fluid model (TFM), the discrete element model (DEM), and the multiphase particle-in-cell (MP-PIC) model for modeling a wide range of multiphase flows [17]. MFiX has been widely used for more than three decades and is well validated, particularly for gas-solid fluidized beds, by TFM [18, 19], DEM [20-22] and MP-PIC models [23-25]. In industrial-scale semi-dense multiphase flows, where the focus is more on trends than solution precision, the MP-PIC model is best suited because it uses a statistical averaging technique that enables simulations to be quickly advanced with minimal particle-level overhead [17]. In addition, using solid-phase normal stress for particle interactions rather than direct consideration of particle collisions can remarkably accelerate computation speed on CPU–GPU hybrid computing platforms [26].

The present study aims to develop an equation-based parallelization method toward efficient and effective GPU acceleration. The TF-based platform allows independent computations for each transport equation by assigning variables into separate processor devices. This enables memory- and computation-efficient, intuitive parallelization of solution procedures depending on the computation load for each equation. The developed MFiX-AI code has been rigorously verified by comparing the number of matching digits in the results, and a dramatic increase in speed was established by using a single GPU node rather than dozens of multicore parallel CPU nodes. Furthermore, MFiX-AI with multiple GPUs can natively deploy arbitrary artificial intelligence and machine learning (AI/ML) models at any point within the CFD code, which allows further optimization in performance.

## 2  Methodology

### 2.1  MP-PIC model

The MP-PIC model in MFiX consists of a Eulerian model for continuous fluid phase and a Lagrangian model to track the position and trajectory of solid particles. Instead of resolving each individual particle, MP-PIC uses the concept of "parcels," wherein groups of identical particles are represented by parcels that are tracked in a Lagrangian manner. A parcel's volume is defined by $V_{\text{parcel}} = \omega V_{\text{particle}}$ where $\omega$ is the statistical weight, representing the average number of particles per parcel. Several assumptions are involved in the MFiX MP-PIC model [17], for example, particles within a computational parcel are assumed to be spherical in shape. Also, instead of considering particle collisions or using any Newtonian mechanics to calculate the individual particle displacement and velocity, the MP-PIC model creates an aggregated solids stress momentum source term that directly affects local solids velocity. Furthermore, the parcels are assumed to maintain a mean density and not experience any rotation. With these assumptions, the three-dimensional gas phase continuity and momentum equations can be written as follows [23]:

$$\frac{\partial}{\partial t}(\varepsilon_G \rho_G) + \nabla \cdot (\varepsilon_G \rho_G \mathbf{u}_G) = 0, \tag{1}$$

$$\frac{\partial}{\partial t}(\varepsilon_G \rho_G \mathbf{u}_G) + \nabla \cdot (\varepsilon_G \rho_G \mathbf{u}_G \mathbf{u}_G) = -\varepsilon_G \nabla p + \nabla \cdot \bar{\bar{\tau}}_G - \mathbf{F} + \varepsilon_G \rho_G \mathbf{g}. \tag{2}$$

and the equations for parcel motion can be represented as

$$\frac{d\mathbf{x}_P}{dt} = \mathbf{u}_P \tag{3}$$

$$\varepsilon_P \rho_P \frac{d\mathbf{u}_P}{dt} = -\varepsilon_P \nabla p + \mathbf{F}_P + \varepsilon_P \rho_P \mathbf{g} - \nabla \tau_P \tag{4}$$

where $\varepsilon_G$ and $\varepsilon_P$ are the volume fraction of gas and parcel, $\rho_G$ and $\rho_P$ are gas density and parcel density, $\mathbf{u}_G$ and $\mathbf{u}_P$ are three-dimensional velocity vectors for gas and parcel, respectively. $\mathbf{g}$ denotes the gravity vector. $\bar{\bar{\tau}}_G$ in Eq. (2) is the fluid shear stress tensor and $\tau_P$ in Eq.(4) is a frictional stress term proposed by Snider [27] as

$$\tau_P = \frac{p_s \varepsilon_P^\beta}{\max\left[\varepsilon_{CP} - \varepsilon_P, \alpha(1-\varepsilon_P)\right]} \tag{5}$$

Here, $\varepsilon_{CP}$ is the close pack volume fraction, and $\alpha$, $\beta$ and $p_s$ are scalar model parameters. The fluid-particle momentum exchange is represented by the interphase drag force terms $\mathbf{F}$ and $\mathbf{F}_P$ in Eqs. (2) and (4), respectively, where $\mathbf{F}$ is the interpolated force from the parcel location to the corresponding fluid cell, and $\mathbf{F}_P$ is the force at the parcel position. The Syamlal-O'Brien model [28] was employed among the various types of drag models tested to form the force terms. Further discussions on drag models, numerical scheme and discretization are beyond the scope of this study, and interested readers may refer to the MFiX Theory Guide [17] for more details.

## 2.2 Code structure

This section introduces the coupling strategy to establish MFiX-AI and its basic code structure. This study is based on MFiX version 19.3.1 (available at https://mfix.netl.doe.gov/doc/mfix/19.3.1/about.html) and TF version 1 or 2 (with TF 2.x behavior deactivated in the latter).

### 2.2.1 Coupling strategy

In MFiX-AI, the core solution procedure occupying the most computation time was rewritten using TF, while miscellaneous functions such as computational domain handling, initial particle population and post-processing remained in the basic Fortran of classic MFiX. Figure 1 demonstrates the workflow within MFiX-AI. The mfix.f file is placed on the top level of the code hierarchy in MFiX. This file controls major code behavior such as initialization, time marching, and calls to fluid and PIC solvers. A flag 'USE_TF_SOLVERS' was introduced to activate the use of the tf_solver.f file for MFiX-AI instead of following the classic MFiX workflow. There are two calls – the first call INITIALIZE_TF is used to initialize the TF graph, and the second call SOLVE_TF is used to take a separate time step inside the TF graph. The INITIALIZE_TF subroutine in the tf_solver.f file gathers required model parameters and truth tables defined during MFiX pre-processing and passes them into the MFiX-AI code. The data exchange between the Fortran-based MFiX code and the python-based MFiX-AI code is established by the additional C wrapper code tfpywrapper.c. The wrapper code is used to directly transfer memory from Fortran to TF. The python functions are defined in the tf_loading_nn.py file that feeds the placeholders in MFiX-AI and runs the TF graph. The body of the CFD computations are handled within the mfixTF_mod.py file and associated modules. For visualization and data recall, the field variables of interest are transferred back to the Fortran code through the same call chain in the reverse direction. Aside from initialization and data recall, no memory transfers into or out of TF during the solution procedure except the time step value. Further, the entire graph is GPU compatible and can be executed with no host/GPU data transfer except for the desired time step at the head of every time step.

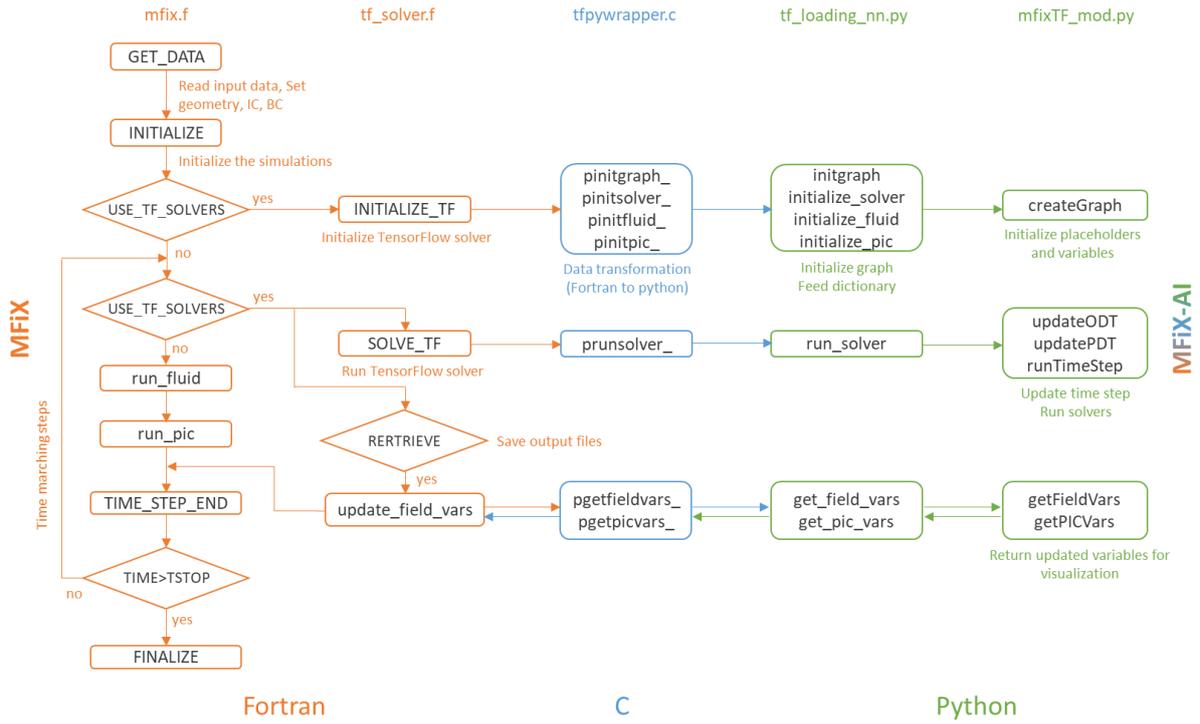

Figure 1: Code structure of the data exchange process between MFiX and MFiX-AI.

### 2.2.2 Parallel computing strategy

MFiX-AI is a direct port from Fortran to the TF Python ecosystem and was designed to run on GPUs. As such, the solution procedure used in MFiX was rewritten using graph representation in TF. MFiX-AI employs placeholders, tensors, and variables supported by TF. Placeholders are used for data input, tensors are immutable data structures within graphs, and variables are mutable data structures. The main part of MFiX-AI is initialized by using placeholders that store values in variables for later recall. A TF variable is designed to maintain a shared, persistent state manipulated by a program [14], so the variables in MFiX-AI are used for the data that should remain on the device between session run calls. Tensors are immutable data structures used for intermediate data that does not need to persist between session run calls. Initialization flows from placeholders through tensors and ends in variable assignment. For calculation of time steps, the graph flows from reads of variables through tensor calculations to variable updates. Because it is only safe to update variables once per session run call, it was decided to create a graph that does a single complete time step and to make repeated calls into TF to advance the time from the Fortran base code. This method also allows for easy matrix construction and solving without the need for repeated field variable and/or matrix data transfers between host and device.

An equation-based parallelization scheme that can utilize multiple GPUs was developed. First, required variables for each component of the momentum equations were declared independently on individual devices (referred to as Udev, Vdev and Wdev), which allowed simultaneous solutions of the momentum equations. If other transport equations such as species and energy are needed, they can also be solved simultaneously on different devices. Next, a set of devices were utilized to solve for pressure, which is a dependent variable dictated by the momentum equation. In the current version of the code, a multi-GPU linear solver feature can be used for pressure. A PICdev was assigned to calculate MP-PIC particle motion and flow. For simplicity the PICdev is always GPU 1 and Pdev for pressure is also GPU 1 when the multi-GPU linear solver mode is disabled. Currently, the PIC equations are solved over a single GPU, which creates a serial workload. Future versions of the code will distribute this across all available devices.

Figure 2 shows schematic diagrams of parallel computing strategies in MFiX and MFiX-AI. The parallel computing strategy for multicore CPU computations in MFiX is a typical MPI-based approach that requires domain decomposition and significant MPI communications to transfer data among the subdomains. The need for MPI communications is often a bottleneck in HPC. In particular, the latencies to communicate dot products often limit the minimum iteration time. Several dot product updates are required at every iteration of the linear solver for every SIMPLE solver iteration. Frequent small communications such as dot products are not advantageous as many studies have shown that as much as 80% of the computation time in modern HPC codes is spent waiting for data to arrive [29-31]. This creates a situation where time to solution does not decrease as processor workload decreases past a certain point, as is seen in most strong scaling curves. The situation can be even more complicated with GPU based solvers because the device efficiency also depends on having enough workload to balance the processing relative to the device global memory access. Even without latency limitations, performance does not scale to small workloads on GPUs, which limits minimum time to solution in parallel workloads as well.

MFiX-AI utilizes an equation-based parallel computing strategy. This kind of strategy is ideal for modern GPUs nodes like the DGX A100 with the NVLink switch. This strategy requires very infrequent communication of entire field variables and no dot product synchronization points in the linear solver unless the parallel pressure solver is turned on. This software architecture trades very frequent small data transfers for infrequent large data transfers. Modern GPU nodes have exceptionally high inter-device bandwidths of up to 7.2 Terabyte/s for NVLink version 3.0, which makes this software architecture tractable since the time to move large data sets between GPUs on the same node is minimal. The inter-device data transfer rates are over 144 times the fastest 400 Gb/s inter-node transfer rates. While this is a substantial increase in bandwidth, latencies have not improved much and are still on the order of 10-30 microseconds. Thus, the equation-based parallelization methodology is expected to work very well on problem scales that can fit in the memory space of a single node. MFiX AI also has an advantage because formation and solve performed on device and host/device data transfers are all but eliminated. The equation-based parallelization strategy can also be more efficient than the current block partitioning in MFiX, where load imbalance can severely affect parallel efficiency. Finally, the equation-based strategy is expected to continue to pay dividends in future HPC systems where GPUs become larger with more memory, as inter-device bandwidth grows significantly (especially relative to inter-node bandwidth) and communication latencies only make marginal improvements.

The parallelization strategy for the pressure equation works well provided that the work per GPU is high and that enough iterations are needed for convergence to overcome the setup overhead. However, the synchronization latencies and workload utilization efficiencies are expected to limit how many GPUs can effectively be used at a given problem scale. This strategy could be extended to transport equations, but these equations usually converge in just a few iterations and it is difficult to make up the costs of setting up the solver for parallel work.

Depending on the available number of GPUs, users can specify GPUs for each Udev, Vdev, Wdev, and Pdev. PICdev is always on GPU 1 in the current version of the code. For example, when only a single GPU is available, all equations are assigned to one GPU, which can be symbolized as 111[1] representing UVW[P] device where PICdev is always assigned to 1. Figure 2 (b) shows an example case of a parallel run with four GPUs (234[1234] decomposition) where three GPUs are assigned to the three velocity components and one GPU is assigned to PICdev. The four numbers in the square bracket denote the workload for pressure distributed to the four GPUs. This configuration is currently the most memory and computationally efficient way to run the code.

For fluid-particle coupling, PICdev is used at least twice in the solution procedure of both the codes depending on whether implicit or explicit coupling is used. At the head of the SIMPLE iteration, coupling from particle to fluid is achieved where the momentum exchange terms **F** in Eq. (2) are calculated. In explicit coupling (PIC expl. in Figure 2), the particle to fluid terms are calculated before the first SIMPLE iteration and are not updated during subsequent SIMPLE iterations. In implicit coupling (PIC impl. In Figure 2), those terms are updated at the beginning of every SIMPLE iteration. Thus, calculation times, which are compared in the performance analysis in Section 4, depend on the coupling method. The second PICdev utilization comes after the SIMPLE iteration and is used to update parcel positions and velocities. The interpolation between particle positions and Eulerian grid values is a known area for further parallel development. The computations could be spread over multiple devices using the high bandwidth available between GPUs on the same node, much like the fluid solver.

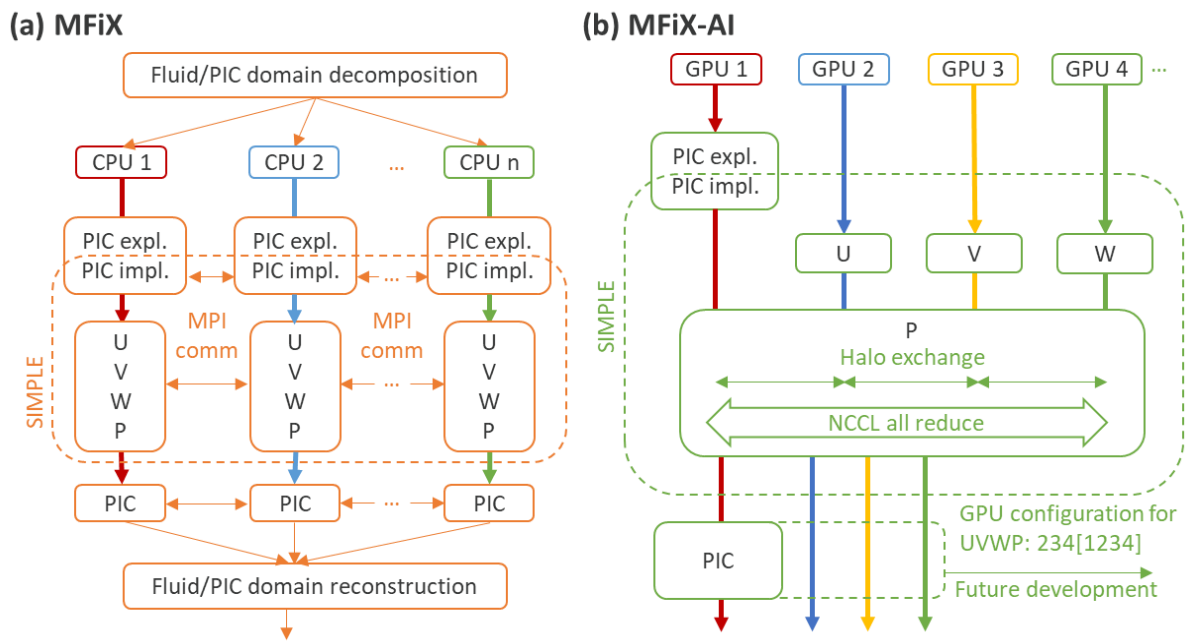

Figure 2: Parallel computing strategy for MFiX on multicore CPUs and MFiX-AI on multi-GPUs. (a) Typical MPI process for the decomposed computational domain used in MFiX, (b) multi-GPU data communication used in MFiX-AI in the TF graph.

*2.2.3   Use of TF graph and custom operator*

MFiX-AI uses TF in graph execution mode, where an efficient computational graph is created before carrying out the actual calculations. A computational graph represents a series of prearranged TF operations that manipulate tensors or tensor-like objects. MFiX-AI employs rank-1 tensors for variables used in calculations and for indices indicating the positions of the computation cells corresponding to the variables. Instead of using element-wise looping, a combination of merge, gather, and scatter operations is used for vector arithmetic. To avoid conditional statements, truth-tables are predefined in Fortran, fed into the graph as rank-1 tensors, and converted to compressed int64 representations to save memory. Although most calculations in MFiX can be done with combinations of standard TF operations, a few complex calculations containing many irregular conditional statements and performance-critical operations are handled through custom operators. Custom operators (or custom ops) in TF allow users to develop functionality that is not defined as default TF operators. The custom ops are written in C++ and whatever language is compatible with other supported devices (such as CUDA for NVIDIA devices and HIP for AMD). The use of custom ops can also be beneficial for reducing the number of kernel calls, optimizing memory usage during the operations, and optimizing memory access patterns.

# 3   Code verification

MFiX-AI was verified by conducting a code-to-code comparison against the classic MFiX code. Figure 3 shows the schematics of the verification case. The test case was an isothermal simulation of a fluidized bed in a rectangular cuboid geometry with dimensions of 12 cm, 72 cm, 12 cm in x, y, and z direction, respectively. The computational domain was discretized using a structured mesh with 27, 162 and 27 cells in x, y, z direction, respectively, such that the total number of cells is 137,924. The constant density and viscosity of the gas phase were set to 1.093 kg/m$^3$ and $1.9 \times 10^{-5}$ Pa·s, respectively. Gravity acts in the -$y$ direction. The domain inlet was set at the bottom ($y = 0$), with gas flow velocity ranging between 0.3 to 1.0 m/s, and the outlet was at the top ($y = 72$ cm). The vertical planes were treated as no-slip walls. The particles were assumed to be of identical size with diameter of 400 μm and density of 2,000 kg/m$^3$. The initial bed height was set to 12 cm with the gas volume fraction set to 0.42 for the bed area. With a statistical weight of 10 (the number of particles per parcel), the total number of parcels present in the domain was 2,983,447. The initial time step was set to 1 ms and varied depending on the convergence of SIMPLE iterations with the given tolerance and maximum number of iterations. The average Courant number according to the initial time step is in the range of $6.75\times10^{-4}$ to $2.25\times10^{-3}$ depending on the gas flow velocity. No preconditioners were used for the linear solvers in either of the codes to ensure that the solver settings are identical.

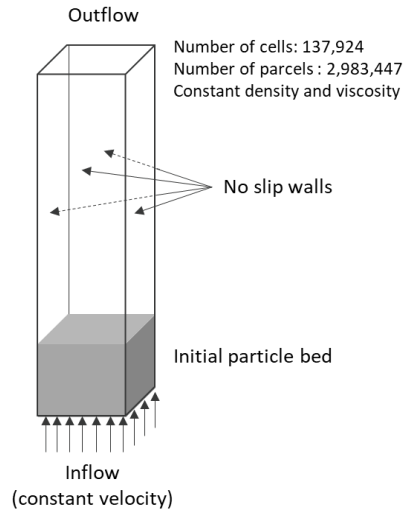

Figure 3: Computational domain and boundary conditions for the verification case.

## 3.1  Verification for a single time step

The test case was conducted using both codes in 64-bit precision to maintain conformity between the MFiX and MFiX-AI simulations. Usually, the initial positions of the particles are set in a randomized manner that precludes an exact comparison of the time evolution of the two codes' simulations. Hence, both simulations were restarted from an MFiX calculation where the bubbling particle motion was active. Residuals of the velocity components and pressure calculated from both codes were in excellent agreement as presented in Figure 4, and a converged solution was obtained in 33 SIMPLE iterations for both codes with the same tolerance for the SIMPLE iterations. In addition, both codes were modified to write the same field variables to file for every cell and every particle at the same point in the solution process. This allowed a comparison of the number of matching significant digits by using the following equation for the field variable $\phi$ from each code.

$$\text{Number of digits matching} = -\log_{10} \frac{|\phi_{\text{MFiX}} - \phi_{\text{MFiX-AI}}|}{|\phi_{\text{MFiX}}|} \qquad (6)$$

This is a very restrictive comparison method as the error is relative to the order of magnitude, and not the absolute difference.

Figure 5 displays histograms for the matching number of digits for the selected variables from both codes at different steps in the solution procedure. The matching number of digits for the velocity components and pressure at the first SIMPLE iteration are shown in Figure 5 (a). The velocity components match within 10 digits and most cells have 14 matching significant digits. However, the pressure matches with fewer digits of precision as it is a quantity derived from the velocity components. In MFiX-AI, the pressure used in the SIMPLE iterations is gauge pressure while absolute pressure was used in MFiX, which allows MFiX-AI to maintain 5 more digits of precision than MFiX. Also, the linear equations for pressure generally take many more iterations to converge than those of the velocity components, which may contribute to error accumulation. Hence, the corrected velocity field at the end of the first SIMPLE iteration (that includes contributions from the predicted pressure field) also shows fewer matching significant digits (similar to that for pressure). That said, most cells match within nine significant digits and the worst cell matched with four. Even with the worst number of digits matching, they are still in very good agreement as it represents less than 0.01% relative difference, given that the two simulations were carried out on different hardware. Note that the pressure values used in Figure 5 (a-b) represent absolute pressure.

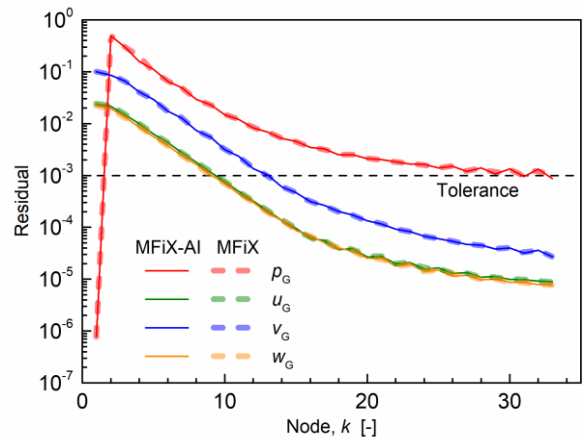

Figure 4: Residuals plots of velocity and pressure from MFiX-AI (solid lines) and MFiX (dashed lines) during SIMPLE iterations at the first time step.

Figure 5 (c) shows that the variables related to the fluid-particle momentum transfer can also influence error propagation. The solid volume fraction $\varepsilon_P$ is calculated using the bilinear numerical interpolation scheme based on the given particle positions from MFiX, but it only matches within eight significant digits. The solids shear stress, $\tau_P$, lost one more matching significant digit after Eq. (5) presumably due to the power function of $\varepsilon_P$ in the numerator. $F_x$, $F_y$ and $F_z$ represent the components of the particle force vector **F** in Eq. (2). Although the particle-related variables in Figure 5 (c) did not significantly affect the gas velocity components as shown in Figure 5 (a), those errors further propagated into the MP-PIC model's calculation of solid motion. Another significant divergence in behavior was observed during the treatment of parcel reflections at the boundary. The problem stems from hard conditionals that determine the occurrence of parcel reflections, which are affected by very small differences in position. A bit perfect match is needed to ensure the same conditional execution in both the codes, and this may not be possible because the computations were performed on different hardware. This unavoidable difference resulted in negative values in Figure 5 (d). Apart from the discrepancies observed on a few particles, the velocity of most particles matched within 11 significant digits and the position of most particles matched within 14 digits, while in the worst case, they matched within four and six digits, respectively.

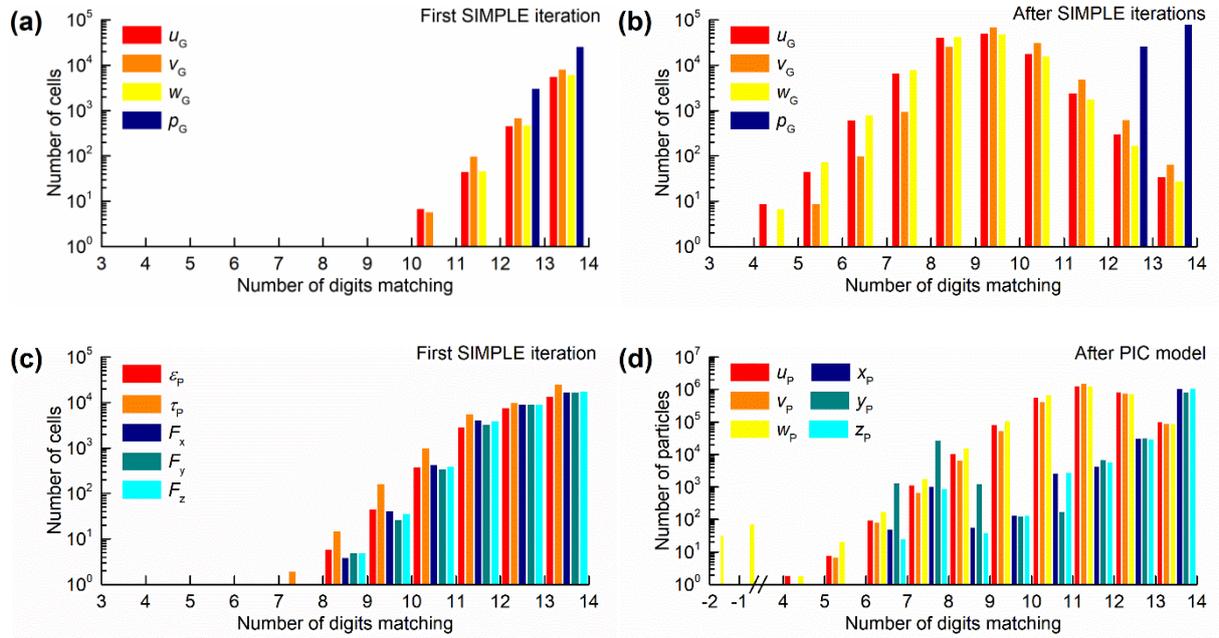

Figure 5: Histograms for the number of matching digits for non-zero quantities obtained from MFiX-AI and MFiX. (a) Comparison of velocity and pressure at the first SIMPLE iteration, (b) at the first time step, (c) variables affecting fluid-particle momentum exchange at the first SIMPLE iteration, and (d) particle position and velocity after PIC calculation.

## 3.2 Validation of bed characteristics

Despite very good agreement observed at the first timestep, the accumulation of bitwise differences leads both results to quickly diverge after a few time steps. This is not a concern as results are expected to be non-deterministic; however, making a side-by-side comparison at every time step is impossible. Therefore, instead of instantaneous differences between the predicted variables, the physical behavior of the fluidized bed was time averaged at a variety of gas inlet velocities. To obtain sufficiently time-independent mean fields, the simulations were run for 102 seconds (in physical time) starting from the initial conditions. The first two seconds were discarded to eliminate the influence of the initial conditions on the time averaged field. A Fast Fourier Transform of the time series of gas volume fraction at a point (located at the middle of the initial bed at [x,y,z] = [6 cm, 6 cm, 6 cm]) was carried out to obtain a bubble frequency spectrum.

Figure 6 demonstrates the results of the fluidized bed simulations from MFiX and MFiX-AI. Figure 6 (a-b) compares the instantaneous and mean fields of gas volume fraction and gas velocity magnitude from both codes at two gas inlet velocities: 0.5 and 1.0m/s. As expected, the instantaneous comparisons do not match, but the mean fields from both codes are qualitatively almost identical. Figure 6 (c) compares both mean gas volume fraction and gas velocity magnitude profiles at the center of the domain in $y$ direction. The maximum differences between the two code profiles are only 0.02 for gas volume fraction and 0.15 m/s for gas velocity magnitude. Moreover, Figure 6 (d) shows a similar dominant frequency and bubble frequency spectrum, which indicates that the bubbling phenomena obtained from the two cases are in excellent agreement. Hence, the series of evidence supports that the bubbling flow dynamics from both codes are very similar.

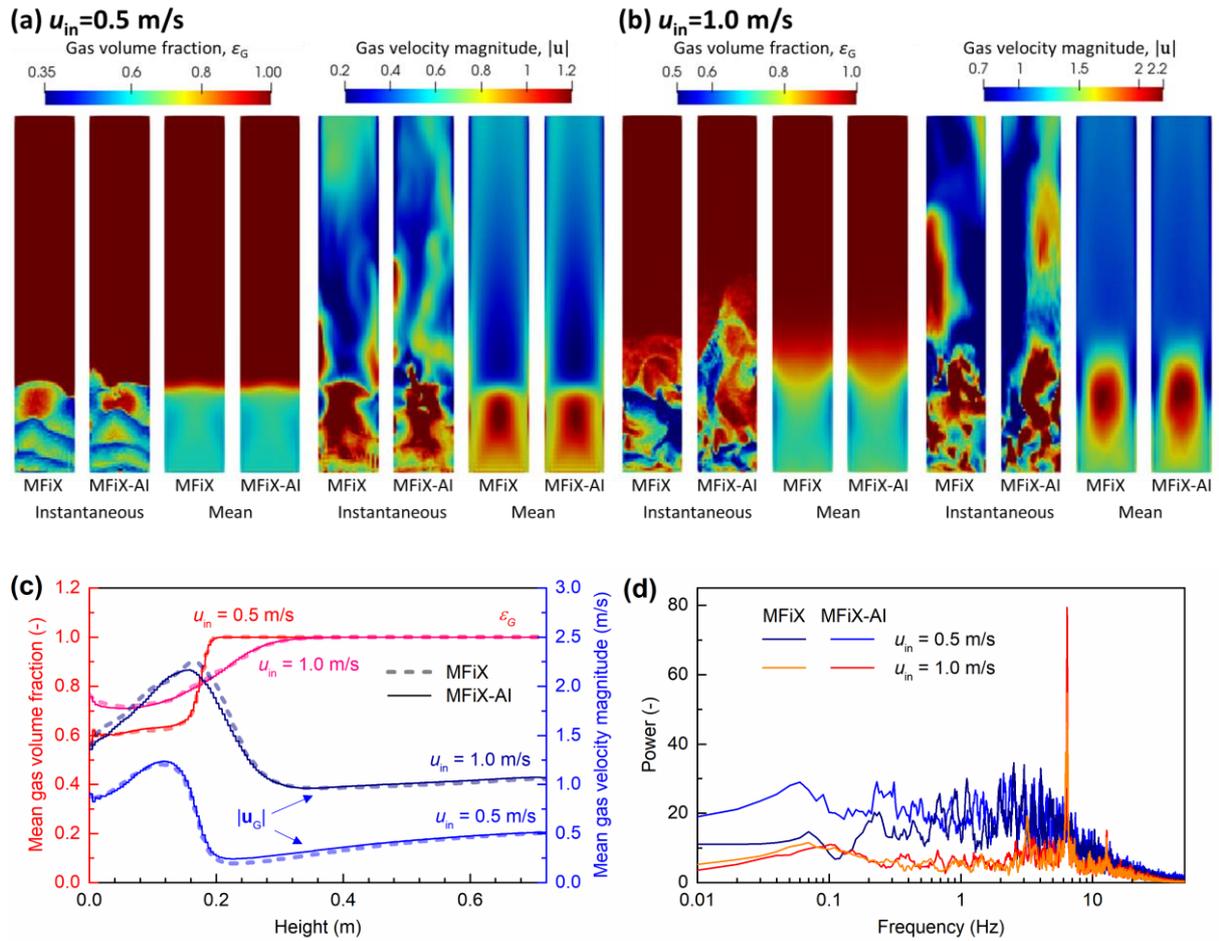

Figure 6: Result of fluidized bed simulations from MFiX and MFiX-AI codes for inlet gas velocity of 0.5 and 1.0 m/s. Instantaneous and time mean fields of gas volume fraction and gas velocity magnitude from both codes for inlet gas velocity of (a) 0.5 m/s and (b) 1.0 m/s. (c) Comparison of mean gas volume fraction and mean gas velocity magnitude profiles at the center of the reactor. (d) Comparison of bubble frequency spectrum from both codes.

In addition, Figure 7 exhibits the pressure drop obtained from both codes for inlet velocity of 0.3-1.0 m/s. The pressure drop obtained from the two codes agree very well, with the relative difference being less than 1%. Those values are also close to the bed pressure representing the total weight of the particle bed. The series of verification and validation tests conducted in this section concludes that MFiX-AI can successfully replicate the predictions from the MFiX code, which is already well validated with physical systems [23-25].

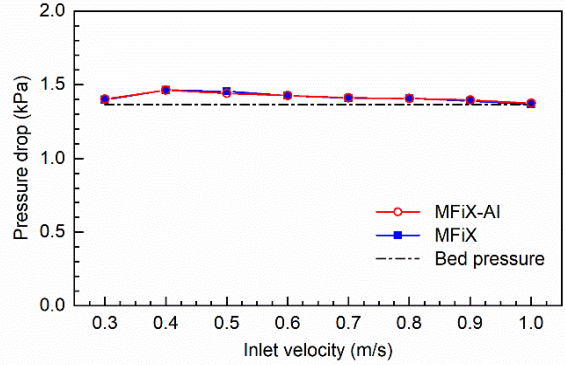

Figure 7: Comparison of pressure drop obtained from the results of both codes running with the gas inlet velocity ranging from 0.3 to 1.0 m/s.

## 4  Parallel scaling performance

The performance enhancement of MFiX-AI against MFiX was evaluated by running identical example cases on Nvidia A100 GPUs (NVLink connected on a DGX A100), each with 80 GB memory for multi-GPU runs (A100-SXM-80GB). The JOULE 2.0 supercomputer at NETL, utilizing Xeon Gold CPU (6148 20C 2.4GHz) with Intel Omni-Path 100Gbit interconnect, was used for the MPI-based multicore-CPU runs. The performance analysis consists of strong scaling analysis with fixed problem size, and weak scaling analysis with problem size fixed per CPU node or GPU. First, single phase calculations without particles were conducted to investigate the speed gains solely for solving the steps in the SIMPLE algorithm. A backward-facing step (BFS) flow case was chosen as the test case for the single-phase calculations. Figure 8 shows the computational domain spanned 9.8 cm, 4.9 cm and 98 cm in $x, y, z$ directions where an internal block with dimensions of 4.9 cm, 4.9 cm, and 9.8 cm was used to construct the step. Fluid viscosity and density were set to $1.8 \times 10^{-5}$ Pa·s and 1.0 kg/m$^3$, respectively, and the inlet velocity (directed along the +z direction) was set to 1 m/s. Secondly, a PIC example case was also used for performance analysis, although the domain was slightly different from that used in the validation study. The computational domain was a cuboid as for the verification case shown in Figure 3. The dimensions are 12 cm, 12 cm, 36 cm in $x, y$ and $z$ directions, and the initial bed height was set to 12 cm. For efficient parallel computation, the longest side was aligned along the $z$ direction, and gravity acts in the -$z$ direction. The density and viscosity were the same as those in the performance analysis study. The inlet gas velocity was set to 0.15 m/s, and the particle diameter and density were specified as 200 μm and 2000 kg/m$^3$, respectively. For both single-phase and PIC examples, the number of cells and particles varied depending on the scaling methods. For both cases, Norm_g, which is a factor to normalize the gas continuity equation residual in MFiX, was set to unity.

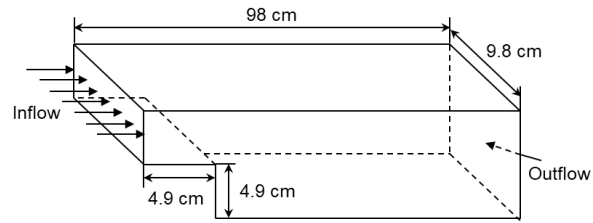

Figure 8: Computational domain and boundary conditions for the BFS benchmark case.

## 4.1 Strong scaling

Strong scaling analyses for single- and multiphase cases were conducted to compare the speed gains and scalability of both MFiX-AI and MFiX codes. Figure 9 (a-b) exhibits the results of strong scaling for single-phase BFS calculations. While the number of cells was fixed as 10,001,880 (10M), 23,708,160 (20M) and 30,802,500 (30M) in this analysis, the number of GPUs used for MFiX-AI calculations varied from one to eight, and the number of nodes for MFiX calculations varied from one to 25, where one node consists of 40 CPU cores. The timestep was initially set to $10^{-4}$ seconds to guarantee convergence from the initial condition, and gradually increased to the maximum timestep of $5\times10^{-4}$ s. The physical time for calculations was 0.5 s. Figure 9 (a) compares the average computation time per SIMPLE iteration of MFiX and MFiX-AI with regards to the size of test problem and the number of CPU nodes or GPUs. MFiX-AI showed a dramatic boost in performance compared to MFiX for all three cases. Figure 9 (b) shows the speed gain of MFiX-AI against MFiX for three combinations of the number of GPUs and the number of CPU nodes. The speed gain generally increased as the computation size increased. The maximum speed gain was 12.4, which corresponds to use of a single GPU node against a single CPU node for the 10M case. It should be noted that calculation times for four GPUs were even slightly faster than the computation times for 25 nodes corresponding to 1,000 MPI ranks on the CPU. Scaling for MFiX AI stopped improving with four GPUs as this problem involves only four equations. If more equations were used, the scaling may continue provided that the parallel solver for pressure can continue to efficiently scale.

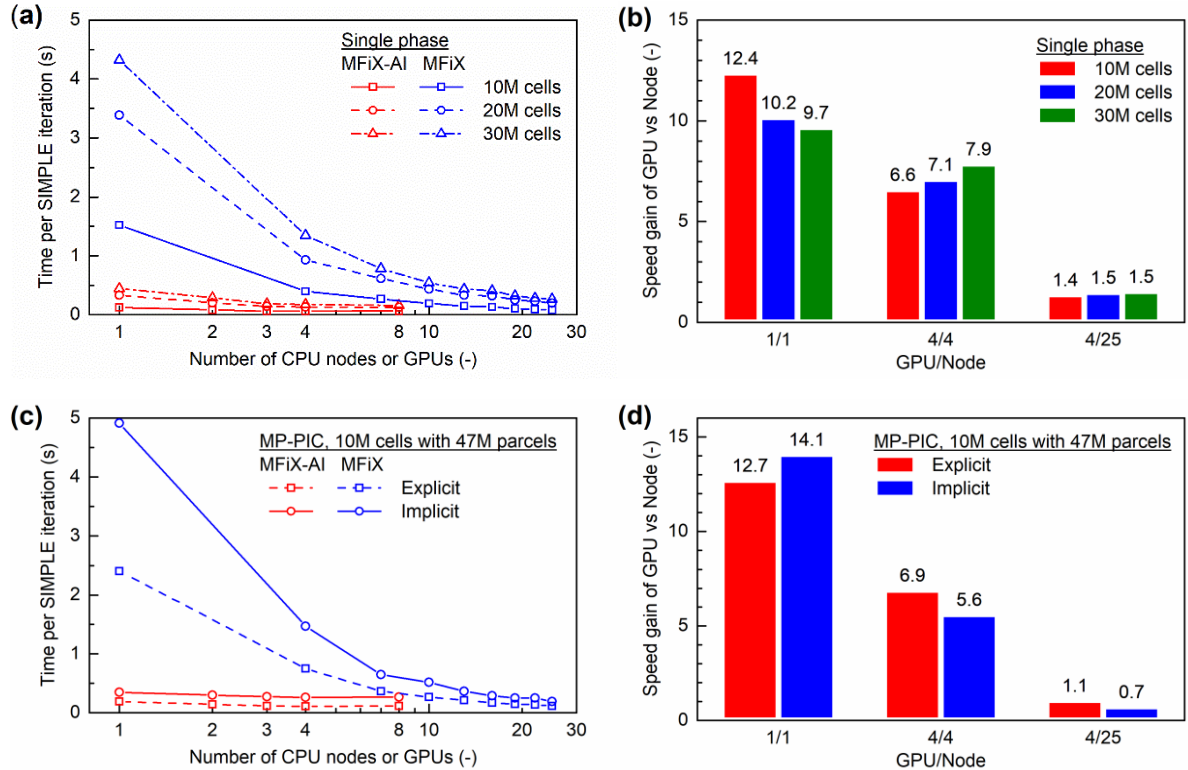

Figure 9: Strong scaling analysis for (a-b) single phase backward facing step flow computations and (c-d) MP-PIC fluidized bed computations obtained by MFiX-AI with multiple A100 GPUs and MFiX with multicore CPU nodes. The average time per SIMPLE iteration of both codes for (a) single phase in terms of cell counts, and (c) those for MP-PIC computations in terms of particle coupling method. Speed gain of MFiX-AI against MFiX for three pairs of GPU/CPU node combinations for (b) single phase computations and (d) MP-PIC computations.

Figure 9 (c-d) shows the results of strong scaling for a benchmark case of PIC calculations with a fixed number of cells (10,125,000; ~10M) and a fixed number of parcels (47,853,435; ~47M). The initial timestep was set to $5 \times 10^{-4}$ seconds, and the 0.5 seconds (physical time) long calculation was used for the strong scaling analysis. The number of GPUs and CPU nodes employed are identical to those used in the single-phase strong scaling analysis. However, instead of comparing the cases with different computation sizes, this comparison focuses on computation times depending on the coupling method. The time per SIMPLE iteration was estimated by dividing the total wall time spent in the SIMPLE routine by the total number of SIMPLE iterations. Thus, the time for PIC computations in both explicit and implicit coupling were included. The difference in calculation times for implicit and explicit coupling is only dictated by the different number of calculations for the momentum exchange terms **F** in Eq.

(2). In explicit coupling, this is calculated once before the start of the SIMPLE loop, while, in implicit coupling, the calculation is performed at the start of every SIMPLE iteration. In Figure 9 (c), the average computation time per SIMPLE iteration for MFiX-AI with a single GPU is much less than MFiX with a single CPU node and is comparable to MFiX with 25 CPU nodes. This dramatic boost in the MFiX-AI computation speed is more clearly shown in Figure 9 (d), which presents the speed gain of MFiX-AI against MFiX. The speed gain of single GPU compared to a single CPU node is approximately 14. The optimal case of MFiX-AI using four GPUs shows excellent performance compared to MFiX using multiple CPU nodes. The computation time for the optimal case is very close to that for MFiX using 25 nodes (=1000 CPU cores); for the explicit case, MFiX-AI is 10% faster than MFiX while for the implicit case MFiX-AI is only 30% slower than MFiX. The observations in both single- and multiphase computations clearly show the dramatic reduction in computation time for MP-PIC calculations using MFiX-AI compared to MFiX. It can be further improved by parallelizing the MP-PIC solver, optimizing the custom operations for hardware-specific implementations, and applying machine learning algorithms for initial guesses of SIMPLE iteration.

## 4.2 Weak scaling of MFiX vs. MFiX-AI with a fixed number of GPUs

A weak scaling analysis for MFiX with a fixed number of cells and/or parcels per number of CPU cores was compared to the MFiX-AI calculations with a fixed number of GPUs for single-phase BFS flow, PIC with a constant number of parcels, and PIC with a variable number of parcels. The test conditions were identical to those used for the strong scaling analysis, with different initial timestep and physical time for calculation, which were $10^{-3}$ seconds, 50 seconds for BFS case and $5 \times 10^{-4}$ seconds, one second for both PIC cases. For MFiX, the cell per node is fixed to approximately 400,000. For PIC with constant parcels, the number of parcels is set to 3,998,062 (4M), and for PIC with variable parcels, the number of parcels per node is set to approximately 4M. For comparison, identical calculations were employed in MFiX-AI but with a constant number of GPUs.

Figure 10 shows the results of the three scaling analyses. In the BFS case, as shown in Figure 10 (a), the time per SIMPLE iteration for MFiX-AI increased almost linearly as the number of cells increased, while for MFiX it showed a gentler increasing trend. The increasing trend for MFiX-AI was expected, given that the workload increased per device. However, the scaling curves for MFiX should be nearly flat in an ideal case. However, almost 2% of the volume was taken out for the step, resulting in a load imbalance, which led to less-than-ideal scaling.

If more than 8M cells are used for MFiX-AI with two GPUs, the computation time will be larger than MFiX for the single-phase calculations (since the MFiX runs were not restricted by the maximum wall time). The MFiX-AI runs with four GPUs were almost consistently faster than MFiX, with a difference in time per iteration of approximately 0.02 seconds (corresponding to total wall time of 1 hour) for all test conditions in this study. It is interesting to note that MFiX-AI with four GPUs scaled almost as well through increased load as MFiX in weak scaling with a slight load imbalance.

Figure 10 (b) displays the weak scaling results for PIC calculations with constant parcels. As the number of cells increases, the MFiX time per iteration dramatically decreases until 3M cells due to the decreasing number of parcels per CPU. The benefit of continued mesh refinement is lost because of particle load balancing in the fluid bed. However, this behavior is not observed with MFiX-AI because the PIC calculations are carried out in a single device (PICdev) and there is no spatial domain decomposition. The computation time for MFiX-AI increases approximately linearly with the cell count. There is almost no difference between computation times of explicit and implicit coupling in MFiX. For MFiX-AI, the computation times for implicit and explicit coupling shifted by an approximately constant value, where the offset represents the additional amount of work in PICdev in SIMPLE iterations. This illustrates the need to parallelize the PIC workloads in future versions of the code. It is noted that MFiX-AI is faster than MFiX for all explicit workloads tested as the increased serial workload for explicit coupling was negligible in performance.

Figure 10 (c) exhibits the same type of scaling analysis for PIC with variable parcels and cells. In this case, the computation times for MFiX-AI show a linear increasing trend, while those for MFiX are almost flat with a slight increase. For explicit coupling, MFiX-AI is always faster than MFiX, while, for implicit coupling, MFiX-AI is only faster than MFiX for less than 7M cells and 70M parcels. The rapidly increasing computation time in the implicitly coupled weak scaling shows the incredible expense involved in interpolating between the Lagrangian particles/parcels and the Eulerian grid, especially in MFiX-AI. This workload scales in proportion to both grid size and particle/parcel count. In general, implicit coupling should not be used unless there is a clear accuracy, need, or advantage for it. Further this illustrates the need to parallelize this workload effectively in future versions of MFiX-AI as the current version of the code only uses one GPU for the PIC portions of the code. That said, the explicit coupling in MFiX-AI is very competitive to the CPU-based MFiX version.

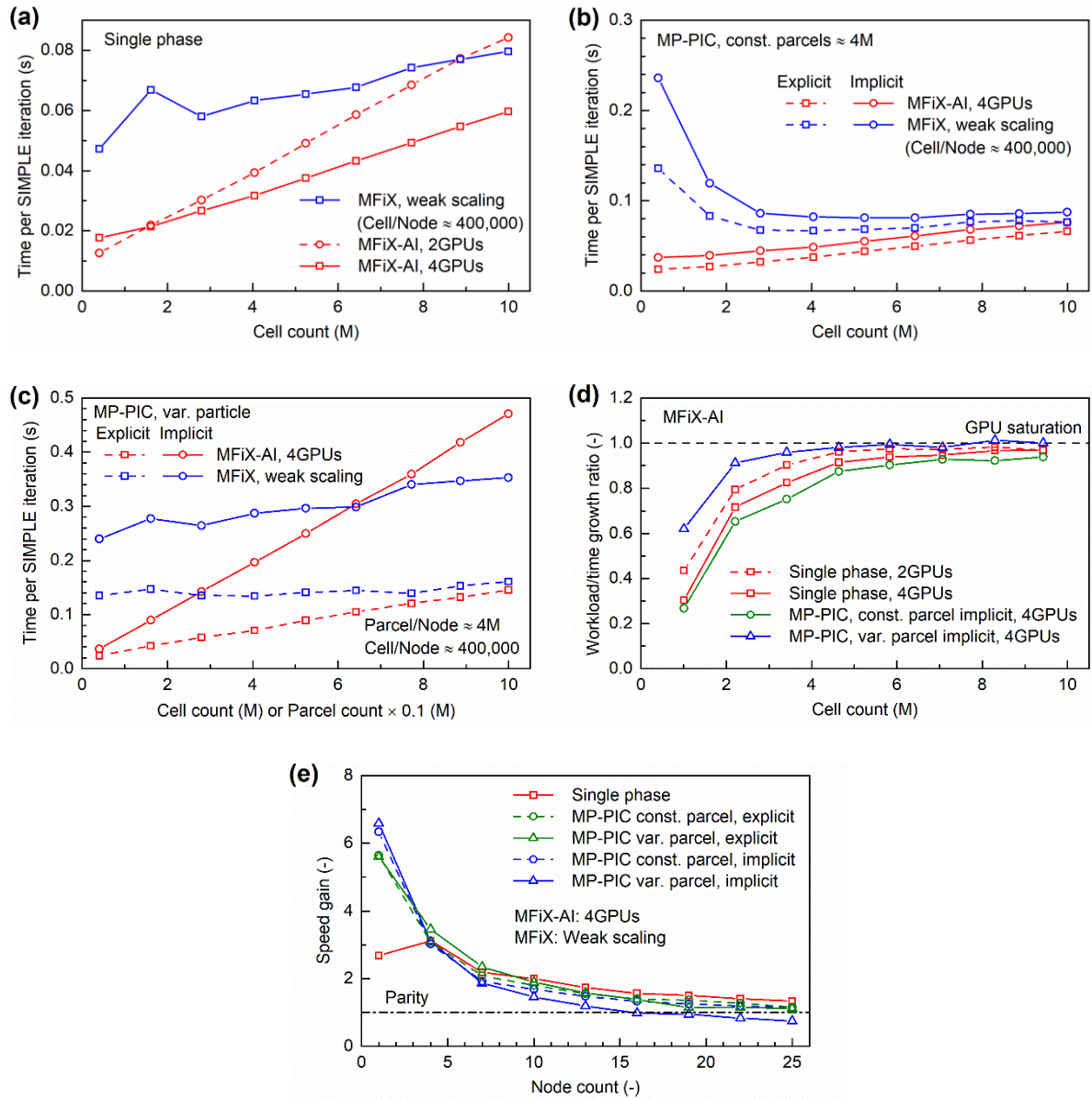

Figure 10: Weak scaling analysis of MFiX for (a) single phase BFS flow, (b) MP-PIC fluidized bed calculation with constant 4 million (M) parcels, and (c) variable parcels with 4M parcels per node. Identical cases of MFiX-AI with fixed number of GPUs were compared. The number of cells is 400,000 per node for all three cases. (d) The ratio of increments of time per SIMPLE iteration and workload for MFiX-AI computations. (e) Comparison of speed gain of MFiX-AI against MFiX for three weak scaling cases.

Figure 10 (d) shows the ratio between time spent for SIMPLE iterations and workload in MFiX-AI computations. The ratios for different cases vary depending on the problem size per number of GPUs employed. A ratio of unity implies GPU workload saturation. In general, workloads become saturated in the 2-5 million cells/parcel range, which is consistent with other reported saturation limits for stencil problems on GPUs [32]. Figure 10 (e) compares the speed gain of MFiX-AI against MFiX for the three weak scaling cases. For the single phase BFS case, the maximum speed gain is approximately three times and it almost linearly decreases with increasing number of nodes for MFiX runs. Including the PIC calculation, the maximum speed gain for constant parcels and variable parcels is approximately seven for both implicit and explicit coupling. Speed gain steeply decreases until 10 nodes and further decreases linearly as the number of nodes increases. It is noted that the PIC calculation of MFiX-AI with constant parcels carried out on a single GPU are even faster than the corresponding calculation in MFiX on 25 nodes. MFiX-AI runs with variable parcels are still beneficial as compared to the corresponding cases in MFiX on less than 15 nodes and almost equivalent to those on 22 and 25 nodes. In most cases, MFiX AI with just four GPUs scales almost as well as weak scaling on CPU nodes with MFiX. The performance in MFiX AI on four GPUs is similar to MFiX runs carried out on many hundreds to thousands of CPU cores.

## 5    Energy consumption efficiency

The dramatic reduction in computation time leads to significant energy savings. Multi-CPU MFiX computations were conducted on the JOULE supercomputer equipped with Intel Omni-Path interconnects. With 64 Omni-Path Edge and two Omni-Path Director switches, each consuming 187 W and 229 W of electricity, respectively, the power consumed by Omni-path interconnects corresponding to each of the 1856 nodes on JOULE is estimated as 6.7 W. Thus, the total power per each node, defined by the sum of the CPU node (800 W) power requirement and the interconnect per node (6.7 W) power requirement, is 806.7 W. Energy consumption of MFiX-AI is estimated from the maximum power requirement of the DGX A100 system, which is 6.5 kW [33]. By using four A100 GPUs (each of which consumes 400W [34]) out of eight GPUs, the estimate total power for running MFiX-AI is 4.9 kW. Figure 11 (a-b) compares the energy consumption of both codes for single SIMPLE iteration in single phase and MP-PIC calculations. In strong scaling analysis, the energy consumption of MFiX-AI with four GPUs is the lowest for the single phase, while slightly increasing from one to four

GPUs for MP-PIC. This evidence consistently shows that the use of four GPUs is optimal. The energy savings was estimated by comparing the energy consumption of four GPUs against that of 25 CPU nodes. Despite the higher power consumption rate of DGX A100 compared to a single CPU node, the MFiX-AI highly benefits from fast computation time. The energy savings of four GPUs against 25 nodes is 82% for the single phase and 67% for the MP-PIC calculations. In weak scaling, the energy consumption of both codes increases monotonically with an increasing cell count. The energy savings for weak scaling analysis is shown to be up to 82% for single phase and 90% for MP-PIC calculations. This study indicates that there are distinct benefits of using multi-GPU computations on a single node relative to as many as 25 CPU nodes. Although it is not straightforward to convert a CPU-based CFD code to a new platform that enables effective GPU acceleration, the benefits in speed gain and energy consumption will be a strong motivating factor to adopt the new paradigm.

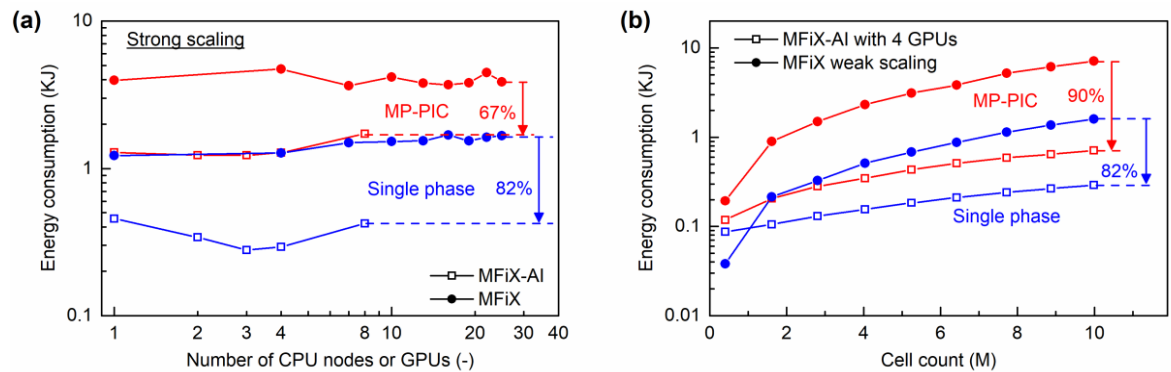

Figure 11: (a) Estimated energy consumption for a single SIMPLE iteration in the strong scaling analysis for single phase with 10M cells and MP-PIC with implicit coupling. (b) The energy consumption of weak scaling analysis for MFiX and MFiX-AI with four GPUs.

# 6 Conclusion and outlook

This study is aimed at developing an equation-based GPU parallelization strategy that enables acceleration of CFD simulations with a multiphase PIC model. For this purpose, MFiX, a well-established and thoroughly validated computer code for multiphase flow, and TF, a powerful open-source library for optimized matrix calculations on CPUs and GPUs, were coupled by replacing the core of a CFD solution procedure in MFiX by a TF-based computer code, MFiX-AI. MFiX-AI distributes the variables into each assigned device, which allows straightforward independent calculations for each transport equation, simultaneously. Thus, an intuitive equation-based parallelization has been accomplished with no domain decomposition. The developed MFiX-AI code produces results that are not bit perfect but match within many significant digits.

The equation-based parallelization strategy within MFiX AI was found to work exceptionally well. Time parities for MFiX AI were found to be above 15 CPU nodes in all cases and were most typically well above 25 nodes. For single-phase cases, MFiX AI with just four GPUs with variable work per device was found to scale almost as well as MFiX in weak scaling at near optimal workload per core. MFiX AI is able to scale out to many tens of millions of cells in single phase. At all tested conditions, MFiX AI was faster than MFiX by a significant margin. Extending the scaling curves shows that MFiX AI is preferrable to MFiX at scales all the way up to approximately 15 million cells. As hardware and code improvements are made, this balance will shift out significantly.

For most multiphase simulations, particle/parcel load imbalances greatly affect the time to solution for domain-decomposed Eulerian Lagrangian software architectures. When particles stack up or are naturally imbalanced, as with a fluid bed simulation, fixed domain decomposition methods struggle to scale well as some computing units get overloaded with more particle calculations than others. Load imbalance is even more severe in complex systems such as recirculating fluidized beds. MFiX AI does not have this issue and shows promise to be an effective method for multiphase calculations in a very similar manner to MFiX. MFiX AI is already a competitive software platform on GPUs for explicit coupling with solids.

# 7 Acknowledgements

This work was supported by Science-based Artificial Intelligence (AI) / Machine Learning (ML) Institute – SAMI, and the project, CFD for Advanced Reactor Design – CARD. We appreciate NVIDIA providing access to NVIDIA Solutions Lab (NSL-B) for computations on DGX A100s. We also thank Dr. Madhava Syamlal for his inspiration and useful comments. Dr. Syamlal provided significant inspiration, motivation, and support for this work. We congratulate Dr. Syamlal on his retirement and wish him the very best on this new chapter of his life. We will never forget all that you have done for us in your long and prosperous career.

# 8  Disclaimer

This project was funded by the Department of Energy, National Energy Technology Laboratory an agency of the United States Government, through an appointment administered by the Oak Ridge Institute for Science and Education. Neither the United States Government nor any agency thereof, nor any of its employees, nor the support contractor, nor any of their employees, makes any warranty, expressor implied, or assumes any legal liability or responsibility for the accuracy, completeness, or usefulness of any information, apparatus, product, or process disclosed, or represents that its use would not infringe privately owned rights. Reference herein to any specific commercial product, process, or service by trade name, trademark, manufacturer, or otherwise does not necessarily constitute or imply its endorsement, recommendation, or favoring by the United States Government or any agency thereof. The views and opinions of authors expressed herein do not necessarily state or reflect those of the United States Government or any agency thereof.

# 9  Code availability

The MFiX-AI code is available as part of the MFiX suite of codes that are provided through NETL's Multiphase Flow Science (MFS) web portal (https://mfix.netl.doe.gov).